\documentclass{article}

\usepackage[labelfont=bf]{caption} % Bold figure number
\usepackage[font={small}]{caption}   %% FIGURE CAPTION FONT MODIFICATION

\usepackage{graphicx}% Include figure files
\usepackage{dcolumn}% Align table columns on decimal point
\usepackage{bm}% bold math
\usepackage[pdftex,colorlinks=true,linkcolor=blue,citecolor=blue]{hyperref}
\usepackage[utf8]{inputenc}
\usepackage{lipsum}  
\usepackage{authblk}
\usepackage{amsfonts}
\usepackage{placeins}
\usepackage[numbers, sort&compress]{natbib}  %%% For compressing the reference numbering

\usepackage[strict]{changepage} % Adjust figure and table width
\usepackage{nicefrac}

\begin{document}

\title{\centering Bound states at partial dislocation defects in multipole higher-order topological insulators}

\author[1]{Sasha S. Yamada}
\author[2]{Tianhe Li}
\author[2]{Mao Lin} 
\author[1]{Christopher W. Peterson}
\author[2]{Taylor L. Hughes}
\author[3*]{Gaurav Bahl}

\affil[1]{\footnotesize{Department of Electrical and Computer Engineering}}
            %, University of Illinois at Urbana-Champaign, Urbana, IL, USA}}
\affil[2]{\footnotesize{Department of Physics and Institute for Condensed Matter Theory}}
            %, University of Illinois at Urbana-Champaign, Urbana, IL, USA}}
\affil[3]{\footnotesize{Department of Mechanical Science and Engineering}}
\affil[ ]{\footnotesize{University of Illinois at Urbana-Champaign, Urbana, IL, USA}}
\affil[*]{\footnotesize{To whom correspondence should be addressed; bahl@illinois.edu}}

%\date{\today}
\date{}
\maketitle

\vspace{12pt}

The bulk-boundary correspondence, which links a bulk topological property of a material to the existence of robust boundary states, is a hallmark of topological insulators \cite{moore2010birth,hasan2010colloquium,qi2011topological}. 
However, in crystalline topological materials the presence of boundary states in the insulating gap is not always necessary since they can be hidden in the bulk energy bands, obscured by boundary artifacts of non-topological origin, or, in the case of higher-order topology, they can be gapped altogether \cite{CornerCharge,Benalcazar2020Bound,CerjanObservation2020,li2020topological,peterson2020FCA, Peterson2021_discl, Liu2021_discl, Roy2020}. 
Crucially, in such systems the interplay between symmetry-protected topology and the corresponding symmetry defects can provide a variety of bulk probes to reveal their topological nature. 
For example, bulk crystallographic defects, such as disclinations and dislocations,  have been shown to bind fractional charges \cite{teo2010topological, jurivcic2012universal, slager2014interplay, teo2017topological, Peterson2021_discl, Liu2021_discl} and/or robust localized bound states \cite{ran2009one, li2018topological, noguchi2019weak, Grinberg2020WTI} in insulators protected by crystalline symmetries. Recently, exotic defects of translation symmetry called partial dislocations have been proposed as a probe of higher-order topology \cite{queiroz2019partial,tuegel2019}. 
However, it is a herculean task to have experimental control over the generation and probing of isolated defects in solid-state systems; hence their use as a bulk probe of topology faces many challenges. Instead, here we show that partial dislocation probes of higher-order topology are ideally suited to the context of engineered materials. Indeed, we present the first observations of partial-dislocation-induced topological modes in 2D and 3D higher-order topological insulators \cite{Benalcazar61} built from circuit-based resonator arrays. While rotational defects (disclinations) have previously been shown to indicate higher-order topology \cite{Peterson2021_discl,Liu2021_discl}, our work provides the first experimental evidence that exotic translation defects (partial dislocations) are bulk topological probes. Furthermore, arrays of partial dislocations are much easier to design than arrays of disclinations; hence applications based on networks of coupled topological bound modes are more amenable to partial dislocation geometries. 

\vspace{12pt}

Topological insulators are generally characterized by quantized topological invariants that are defined with respect to the bulk symmetries of a system  \cite{moore2010birth, hasan2010colloquium,qi2011topological}.
Here we focus on a particular class of topological crystalline insulators (TCIs), that are characterized by electric multipole moments, which are quantized by the crystal symmetries \cite{Watanabe2020Corner,Ren2021Quadrupole}. For example, inversion symmetry quantizes the electric polarization (i.e. dipole moment per unit cell) of an insulator \cite{zak1989}, 
and this manifests as gapless modes and fractional charge (per unit cell) localized at the boundaries of a material \cite{Su1979,zak1989,hughes2011,turner2012}.
The symmetry-enforced quantization generalizes to higher electric multipole moments, e.g. quadrupole and octupole moments, which become quantized in crystals in the presence of any symmetry under which they flip their sign \cite{Benalcazar61}. Quantized quadrupole and octupole insulators are a subset of higher-order topological insulators (HOTIs), and they host gapless states and/or fractional charges not on their surfaces, but at their corners \cite{Benalcazar61,HOTI,peterson2018quad,quadImhof2018,quadSerraGarcia2018,quadMittal2019,Bao2019_oct,quadQi2020,Xue2020_oct,octNi2020}.

Since the topology of multipole HOTIs is protected by crystalline symmetries, one would naturally expect these insulators to be sensitive to crystallographic defects \cite{ teo2010topological, jurivcic2012universal, slager2014interplay,teo2017topological, Benalcazar61,Roy2020, Peterson2021_discl, Liu2021_discl}. 
Single dislocation \cite{slager2014interplay,li2018topological, Miert2018Dislocation, slager2018translational, nayak2019resolving, Grinberg2020WTI, Roy2020} and disclination \cite{benalcazar2014classification, CornerCharge, Li2020Fractional, geier2020bulk, Wang243602, Peterson2021_discl,Liu2021_discl} defects have been investigated as bulk probes of topology for TCIs where they trap bound states and/or fractional charges.
However, conventional full dislocation defects are not useful probes of multipole HOTIs as they will not elicit a topological bound state or fractional charge response (see Fig.~\ref{fig:full_disloc}) \cite{Benalcazar61, teo2017topological}. Additionally, disclinations have some notable shortcomings in experiments: (i) in solid-state systems they cost a large amount of elastic energy and are difficult to isolate, (ii) in engineered materials they are not convenient to implement experimentally as single disclination defects require a massive rearrangement of the global structure of the lattice, (iii) while they may bind fractional charge in 2D multipole insulators, they are not guaranteed to do so in 3D multipole insulators, and (iv) even when exhibiting fractional charge, disclination cores are often not accompanied by mid-gap topological bound states \cite{Peterson2021_discl,Liu2021_discl}, which precludes the use of the most convenient experimental probes, i.e., spectroscopy.

To circumvent these issues, we instead consider partial dislocation defects as bulk probes of e.g. multipole HOTIs. These defects can occur naturally in solid-state systems, and are formed when regions of incomplete unit cells are inserted or removed from the lattice. In general, dislocation defects are characterized by a Burgers vector. The Burgers vector quantifies the difference in the number of translations required to trace a loop around an ordinary point of the lattice compared to tracing a loop around a dislocation core. When a dislocation core replaces the ordinary point, then the original loop that was traced around the ordinary point fails to perfectly close, and the deficit or excess in the displacement is the Burgers vector \cite{teo2017topological,queiroz2019partial}.
Usually this Burgers vector is a lattice translation vector, but since partial dislocations are formed by inserting/removing incomplete unit cells, their Burgers vector is always a fraction of a lattice vector. 

\begin{figure}[hbtp]
    \begin{adjustwidth}{-1in}{-1in}
        \centering
        \includegraphics[width=1.3\textwidth]{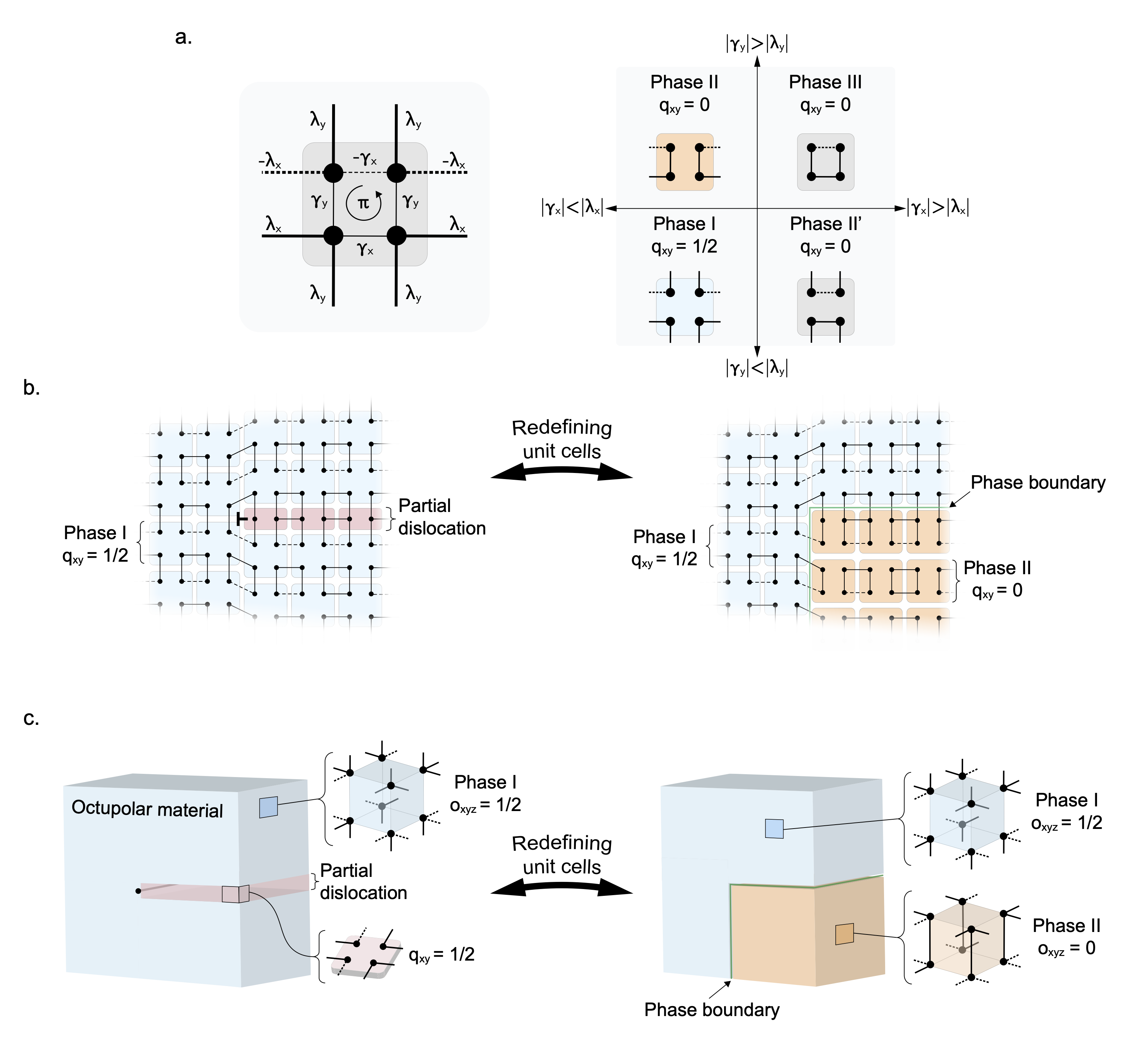}
        \caption{
       \textbf{Origination of topological phase boundaries due to partial dislocations in 2D and 3D multipole HOTI materials.} 
        \textbf{(a)} Possible connections in the minimal model of single quadrupolar TCI unit cell and corresponding topological phase diagram. Intra-cell couplings are labeled as $\gamma_{x,y}$ and inter-cell couplings are labeled as $\lambda_{x,y}$. The dashed lines denote negative coupling. The dimerized limit is illustrated in the phase diagram, so only the dominant connections are shown.
        \textbf{(b)} Equivalence between quadrupolar TCI (blue unit cells) in the topological phase with partial dislocation (red unit cells) and phase boundary between trivial (orange unit cells) and non-trivial (blue unit cells) quadrupolar phases. The dislocation core is denoted as $\boldsymbol\vdash$. The unit cells are depicted in their dimerized limit.  
        \textbf{(c)} Same as (b), but for an octupolar TCI. Here, the partial dislocation is a plane defect and thus terminates along a line within the bulk. The unit cells shown in the inset are depicted in their dimerized limit. 
        }
        \label{fig:cut_glue}
    \end{adjustwidth}
\end{figure}

Let us now illustrate how 2D and 3D multipole HOTIs react to the presence of a partial dislocation. As a specific example, we first consider the 2D quadrupole insulator model from Ref. \cite{Benalcazar61}, which is formed on a square lattice with an alternating weak and strong coupling pattern, and a $\pi$ flux threading each plaquette (see Fig. \ref{fig:cut_glue}a). The resulting model has four degrees of freedom per unit cell, and the $\pi$ flux ensures that the bulk and edge bands are spectrally gapped. 
For different choices of the intra- and inter-cell couplings, this system has four natural, gapped configurations (Fig.~\ref{fig:cut_glue}a), of which only the one having weak intra-cell coupling in both directions has a non-zero, quantized quadrupole moment per unit area of $q_{xy} = \nicefrac{1}{2}$, in units of elementary charge $e$ \cite{Benalcazar61}.\footnote{Strictly speaking, since we are allowing for couplings which break $C_4$ symmetry, but preserve mirror symmetries in $x$ and $y$ this system is a boundary obstructed topological phase.}
There is an interesting subtlety in that, for an infinite or periodic lattice, one can change which configuration has the non-vanishing quadrupole moment just by redefining the choice of unit cell. This is not unique to the quadrupole HOTI, and is also present for the properties of any obstructed atomic insulator \cite{Su1979,Khalaf2021Boundaryobstructed}. Hence, while we cannot give an absolute statement about which configuration has the non-zero quadrupole moment without fixing a unit cell choice, we can distinguish the \emph{differences} between the configurations by making domain walls between them. We will now show that the insertion of a partial dislocation creates exactly such spatial domains of all four configurations, and hence leads to an observable response.

As a frame of reference for the partial dislocation discussion let us fix our unit cell choice so that Phase I is a non-trivial quadrupolar HOTI having $q_{xy}=\nicefrac{1}{2}$. To construct the partial dislocation we now insert a row of partial unit cells into the lattice as in Fig.~{\ref{fig:cut_glue}}b. This partial dislocation is characterized by a fractional Burgers vector ${\bf B}=(\nicefrac{1}{2}) \hat{y}$ in units of lattice constant (left panel of Fig.~\ref{fig:cut_glue}b).
To clearly illustrate the effect of the partial dislocation we can now choose to redefine the unit cells in one quadrant to incorporate the partial dislocation and once again form complete quadrupolar unit cells. We note that this is merely a change of perspective, i.e.,  we are not  modifying the micro-structure of the lattice. For our example in Fig.~\ref{fig:cut_glue}b, we group the partial dislocation with the lower right quadrant of the lattice. 
Compared to the original  unit cells (Phase I in Fig.~\ref{fig:cut_glue}a), the redefined unit cells in the lower right quadrant are in Phase II and no longer host a non-trivial quadrupole moment. This shows us that the partial dislocation has introduced a spatial phase boundary between a trivial and non-trivial quadrupolar HOTI. As a result, we expect fractional charge and a 0D topological bound state to be localized at the higher-order boundaries (i.e., the corner of the phase boundary), which sits at the dislocation core. 

This argument can be similarly extended to 3D octupolar TCIs having partial dislocation planes.
We begin with an octupolar HOTI model of Ref. \cite{Benalcazar61} composed of unit cells having eight degrees of freedom. The configuration where all of the $x,y,z$ inter-cell couplings are stronger than the respective intra-cell couplings hosts a non-trivial octupole moment $o_{xyz} = \nicefrac{1}{2}$, in units of elementary charge $e.$ Now we can insert a partial dislocation by inserting a plane composed of half of the octupole model's unit cells into the original lattice.
The edge of the inserted plane is a partial dislocation line, and the Burgers vector associated with that line is ${\bf B}=(1/2)\hat{z}$ in units of the lattice constant. 
Similar to our illustration for the quadrupolar TCI, we can consider redefined the unit cells that group the partial dislocation with the lower right sector as shown in Fig.~\ref{fig:cut_glue}c. In doing so, we find that the redefined unit cells no longer host a non-trivial octupole moment, and the partial dislocation forms a phase boundary with the original lattice configuration. 
We thus expect 0D topological bound states to be localized at locations where the dislocation line terminates on a surface, or if it is closed, where the dislocation line makes a 90 degree turn to form a corner. 
Although we considered specific models in simple limits, this argument can be generalized to any HOTI that hosts different topology when its unit cells are modified by the addition of a fractional unit cell.

\vspace{12pt}

\begin{figure}  [hbtp]
    \begin{adjustwidth}{-1in}{-1in}
    \centering
    \includegraphics[width=1.4 \columnwidth]{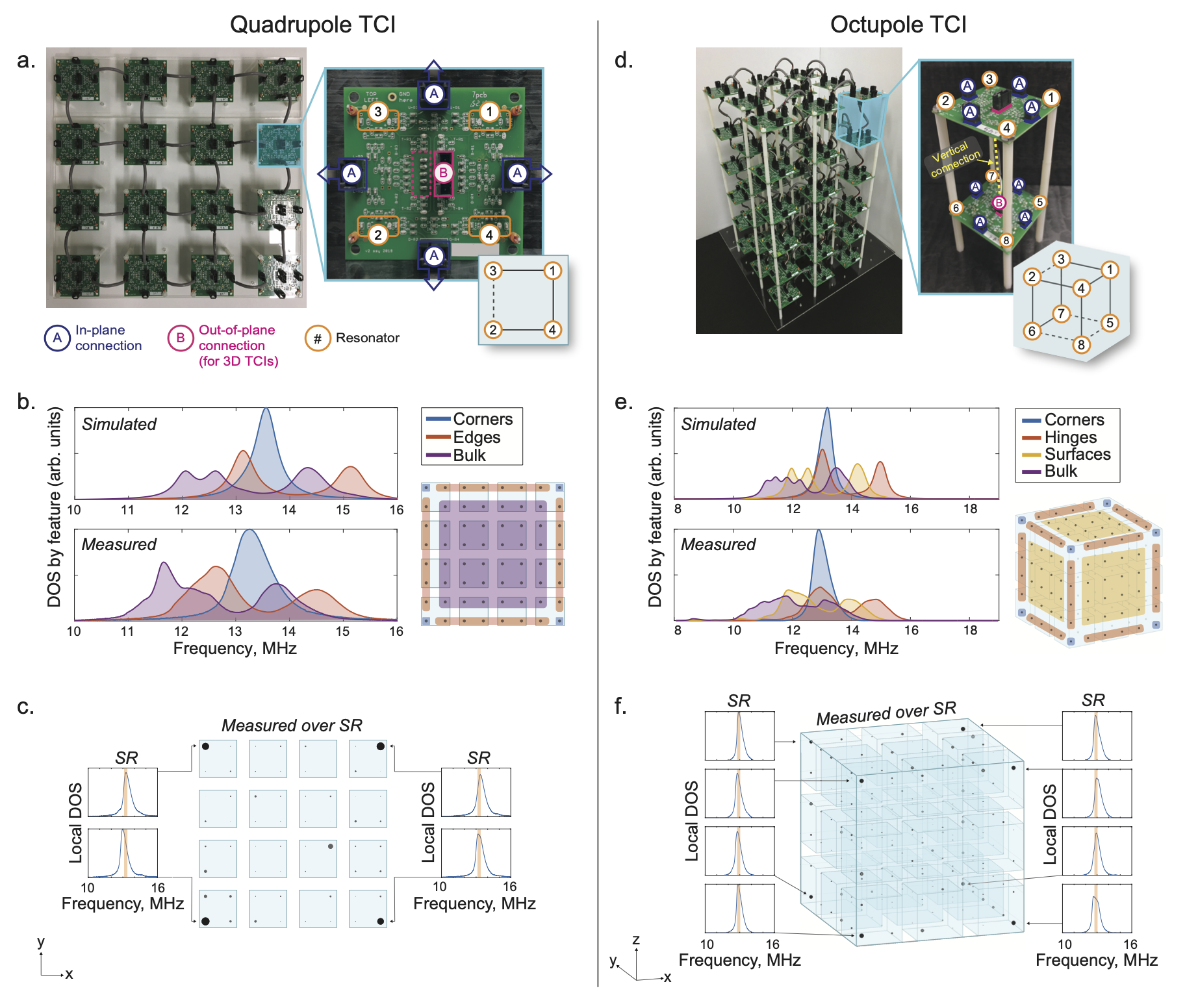}
    \caption{\textbf{Implementation of modular quadrupolar and octupolar TCIs.} 
    Photographs of the \textbf{(a)} quadrupolar and \textbf{(d)} octupolar TCIs, including zoomed-in views of the individual unit cells.
    The inset connectivity diagram represents the tight-binding model within the unit cell, where the dashed lines indicate negative coupling, which implements the $\pi$ flux in each plaquette.
    \textbf{(b)} Measured density of states (DOS) averaged over sites that lie on the corresponding feature groupings for the quadrupolar TCI. 
    \textbf{(e)} Measurements of DOS for the octupole TCI averaged by feature grouping, similar to (b).
    \textbf{(c)} Spatial map of the mode density at each site over the highlighted spectral range (SR: 13.1 - 13.4 MHz), which corresponds to where the corners have the highest mode density and where the edge and bulk spectra are gapped. Each site is represented as a circle with radius proportional to the local mode density. 
    \textbf{(f)} Similarly to (c), we show the mode density at each site for an octupolar TCI excited over the highlighted spectral range (SR: 12.75 MHz - 13.05 MHz). 
    Complete local DOS measurements for all sites in (c) and (f) are presented in Supplementary Figures~\ref{fig:quad_grid1} and \ref{fig:oct_grid1}.
    }
    \label{fig:implementation}
    \end{adjustwidth}
\end{figure}

With an aim to observe these phenomena, we constructed partial dislocations in multipole HOTIs using metamaterials composed of coupled electronic resonators. In this metamaterial approach, each resonator represents a spinless orbital, and the coupling between resonators is analogous to the hopping of electrons between atoms. The metamaterials were assembled using unit cells composed of modular circuit boards, each of which contains a four-resonator plaquette whose connectivity can be fully reconfigured, including the choice of $0$ or $\pi$ flux threading the plaquette. This also allows for arbitrary choice of boundary conditions, e.g., open or periodic boundaries. We consider each individual module as a 4-site unit cell in the quadrupole model, and a pair of stacked modules as an 8-site unit cell in the octupole model as shown in Fig.~\ref{fig:implementation}a,d.
Additional information on the design of the modular circuit boards is presented in Methods. 

We first assembled a 2D quadrupolar TCI with a size of $4\times4$ unit cells (Fig.~\ref{fig:implementation}a), composed of 64 total sites on 16 modular circuit boards.
Radio-frequency (RF) power reflection measurements were performed to obtain the local spectral density of states (DOS) in the metamaterial \cite{peterson2020FCA}. Details on how the RF reflection measurement is mapped to the DOS are presented in Methods. 
Since the bulk quadrupole moment imparts second-order multipole characteristics, we expect spectrally gapless modes localized at the second-order boundaries, i.e., the corners of the 2D system \cite{Benalcazar61, peterson2018quad,quadImhof2018,quadSerraGarcia2018, quadMittal2019,quadQi2020}. The bulk bands and edge bands are expected to be gapped.
These properties were experimentally confirmed in the averaged DOS measurements for the corner, edge, and bulk resonator groups as shown in Fig.~\ref{fig:implementation}b. 
We note that the band gaps of the bulk and edge groups are not spectrally aligned due the breaking of chiral symmetry at the boundaries of the terminated lattice, which leads to a loading effect \cite{peterson2018quad,Peterson2021_discl}.
We further confirm corner localization of the mid-gap modes by spatially mapping the mode density over the entire array at the same frequency as the corner mode (Fig.~\ref{fig:implementation}c). 

We similarly constructed a 3D octupole TCI with a $3\times3\times3$ unit cell configuration, composed of 216 total sites on 54 modular circuit boards, as shown in Fig.~\ref{fig:implementation}d. 
Since the bulk octupole moment imparts third-order multipole characteristics, we anticipate gapless modes at the third-order boundaries, i.e., at the eight corners of a cube \cite{Benalcazar61, Bao2019_oct, Xue2020_oct, octNi2020}. 
The lower-order features corresponding to the bulk, hinge, and surface bands are all expected to be gapped.

This phenomenology was verified by measuring the averaged DOS of the corner, hinge, surface, and bulk groups as shown in Fig.~\ref{fig:implementation}e. Once again, the breaking of chiral symmetry at the boundaries of our terminated lattice causes the band gaps to be spectrally misaligned. 
Corner localization of the gapless modes is again confirmed by spatial mapping of the mode density (Fig.~\ref{fig:implementation}f).

\begin{figure} [hbtp]
    \begin{adjustwidth}{-1in}{-1in}
    \centering
    %\frame{
        \includegraphics[width=1.4\textwidth]{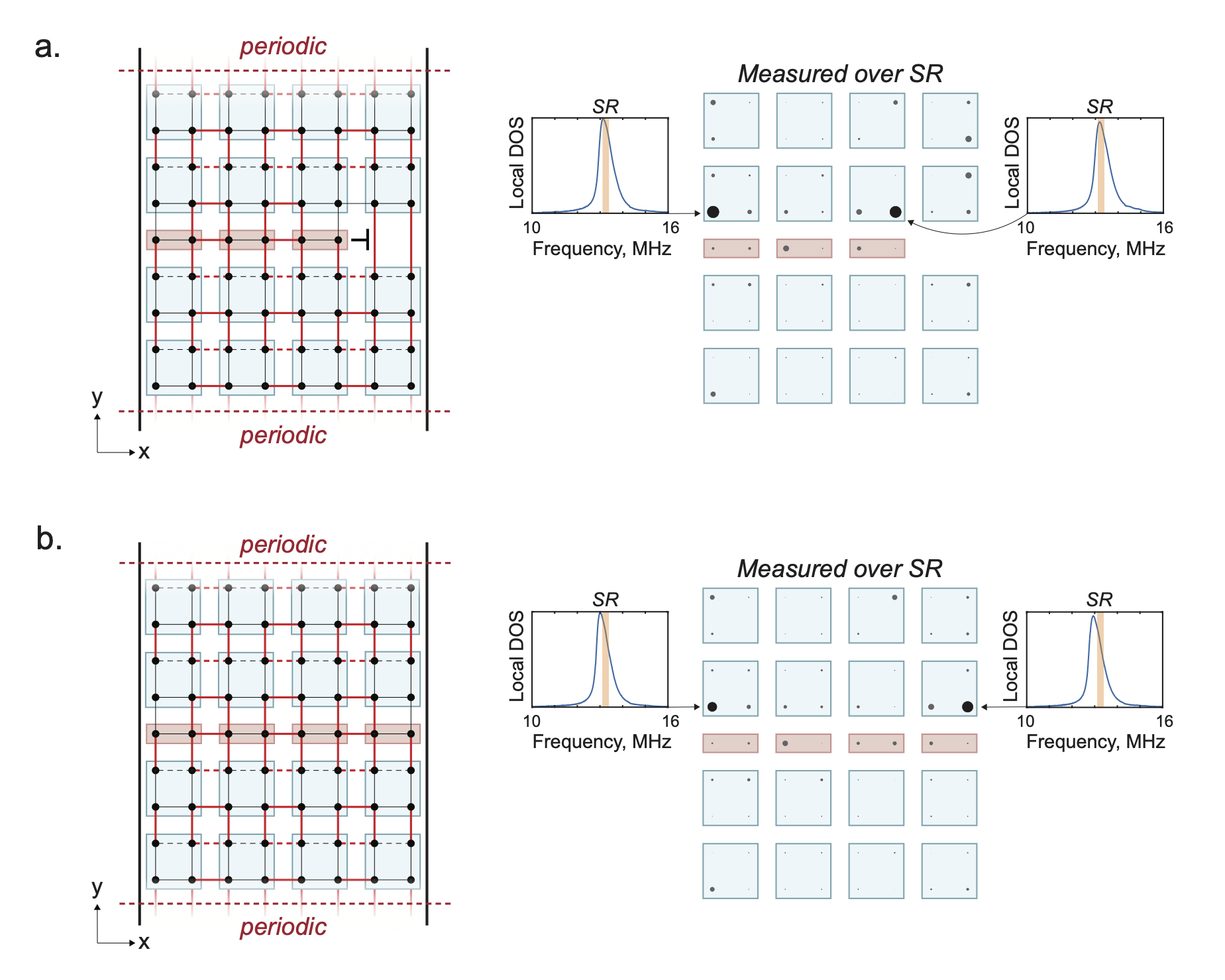}
    %}
    \caption{\textbf{Quadrupolar TCI partial dislocation experiments.} \textbf{(a)} Topology of the measured quadrupolar TCI with an bulk-terminated partial dislocation and spatial map of the mode density at each site over the spectral range (SR) of 13.1 MHz - 13.4 MHz. The blue boxes represent complete quadrupolar unit cells, and the red boxes represent the partial dislocation. The red (black) lines denote strong (weak) coupling, and the dashed lines represent negative coupling. In the spatial map, each site is represented by a circle with radius proportional to the local mode density. The local DOS of the sites with the largest mode densities are shown in the inset and the SR is highlighted in orange.
    \textbf{(b)} Same as (a), but for a boundary-terminated partial dislocation.
    Complete local DOS measurements for all sites in (a) and (b) are presented in Supplementary Figures~\ref{fig:quad_grid2} and \ref{fig:quad_grid3}.
    }
    \label{fig:quad_results}
    \end{adjustwidth}
\end{figure}

\vspace{12pt} 

With the benchmark multipole HOTI characteristics confirmed, we now introduce partial dislocations in these lattices.
In the 2D quadrupolar TCI (Fig.~\ref{fig:quad_results}a), we first add a periodic boundary condition along the y-direction to gap out the outer corner modes and hence minimize boundary effects. 
We then introduce a partial dislocation composed of $3 \times \nicefrac{1}{2}$ quadrupolar unit cells oriented along the x-axis. 
As discussed earlier, we anticipate topological bound modes to be localized near the higher-order boundaries of the phase boundary formed by the inclusion of the partial dislocation. 
However, the exact spatial distribution of these modes depends on how the boundaries of the partial dislocation are coupled to the original lattice. We use this freedom to aid our observations, i.e., by locally deforming the inter-site connectivity at the higher-order boundaries of the partial dislocation we are able to spatially isolate the bound state to only one site (see Fig.~\ref{fig:3Dcoupling}). 
This deformation does not disturb the protective reflection symmetry of the inserted partial dislocation, nor does it destroy the topological phase boundary, so the global properties of the structure remain unchanged. 
Experimental measurements shown in Fig.~\ref{fig:quad_results}a confirm that the bound states are both spectrally gapless and are spatially localized at the higher-order boundaries of the partial dislocation.
Next, we showed that the observed bound states are robust irrespective of whether the partial dislocation terminates within the bulk or at the boundary of the TCI; as might be expected since the boundary itself is gapped.
To achieve this we extend the partial dislocation to form a $4 \times \nicefrac{1}{2}$ unit cell configuration as shown in Fig.~\ref{fig:quad_results}b, where both ends terminate at the boundary. Experimental measurements confirm that the two bound states now simply relocate to the new higher-order boundaries of the partial dislocation, confirming their robustness.

\begin{figure} [hbtp]
    \begin{adjustwidth}{-1in}{-1in}
    \centering
    %\frame{
        \includegraphics[width=1.4\textwidth]{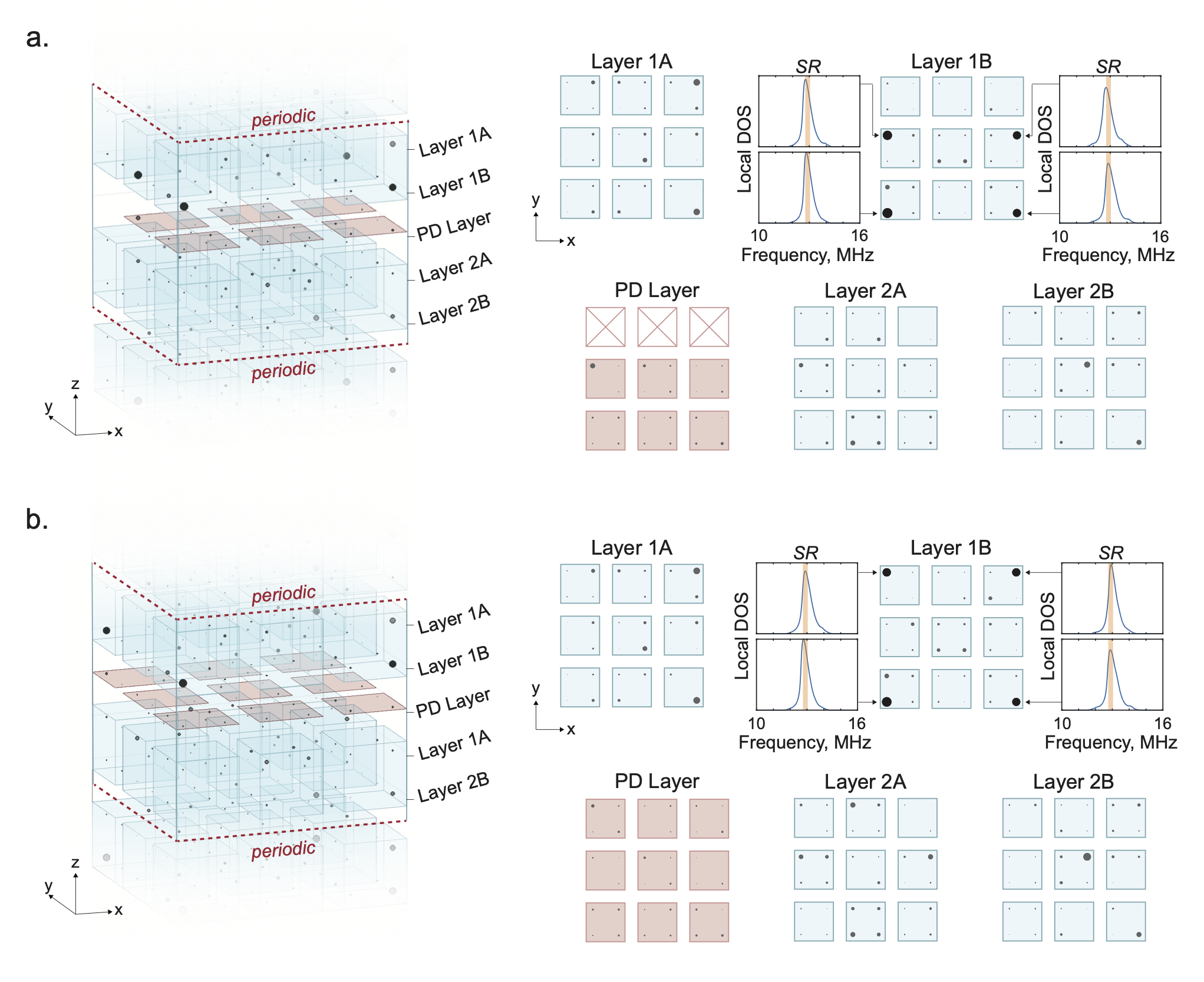}
    %}
    \caption{\textbf{Octupolar TCI partial dislocation experiments.} 
    \textbf{(a)} Spatial map of mode density at each site over the spectral range (SR) of 12.75 MHz - 13.05 MHz in an octupole TCI (blue unit cells) with an bulk-terminated partial dislocation (red unit cells). In the spatial map, each site is represented by a circle with radius proportional to the local mode density over the selected SR. The local DOS is shown in the inset for the sites with the largest mode density, and the SR is highlighted in orange.  
    \textbf{(b)} same as (a), but for a boundary-terminated partial dislocation.
    Complete local DOS measurements for all sites in (a) and (b) are presented in Supplementary Figures~\ref{fig:oct_grid2} and \ref{fig:oct_grid3}.
    }
    \label{fig:oct_results}
    \end{adjustwidth}
\end{figure}

We now move on to investigate partial dislocation defects in the 3D octupolar system.
Similar to the 2D quadrupolar HOTI, we set up periodic boundary conditions along the z-direction to eliminate the outer corner modes and to minimize boundary effects. 
A bulk-terminated partial dislocation was then introduced by changing the middle layer of unit cells into a $3 \times 2 \times \nicefrac{1}{2}$ configuration.
We again locally deformed the inter-site connectivity at the higher-order boundaries of the partial dislocation to isolate the topological bound mode to a single site. Details of the local connectivity are presented in Fig.~\ref{fig:3Dcoupling}.
Once again, the bound states associated with the partial dislocation can be observed as gapless modes at the higher-order boundaries of the partial dislocation.
These results are shown in the spatial map in Fig.~\ref{fig:oct_results}a. 
As before, we confirmed experimentally that these bound states survive when the partial dislocation plane is extended to the outer boundary of the 3D material, using a $3\times3\times\nicefrac{1}{2}$ unit-cell plane. 
These bound states are again shown to be spectrally gapless and localized (Fig.~\ref{fig:oct_results}b).

\vspace{12pt}

In this study, we have demonstrated that partial dislocation defects can successfully probe the topology of TCIs with vanishing dipole polarization, which full dislocations fail to detect. 
Likewise, when compared to disclination defects, partial dislocations have the advantage of simple integration into engineered materials and are a reliable bulk probe of higher-order topology in both 2D and 3D HOTIs.
We also predict that partial dislocation defects will trap fractional charge, as has been shown for disclination defects \cite{Peterson2021_discl, Liu2021_discl}.
Furthermore, we anticipate that the bound states and fractional charges trapped by these defects can be harnessed for potential engineering applications, such as topological lasing, %\cite{TI_laser_theory,TI_laser_exp,TI_laser_1D}, 
as our findings provide a pathway to selectively embed 0D protected states deep within the bulk of 2D or 3D HOTIs
More broadly, partial dislocation defects are abundant in solid-state materials, and may provide a powerful new probe of higher-order topology in quantum materials.
%

%%%%%%%%%%%%%%%%%%%%%%%%%%%%%%%%%%%%%
\section*{Methods}
%%%%%%%%%%%%%%%%%%%%%%%%%%%%%%%%%%%%%

\textit{Design of metamaterial:} 
Since topological crystalline insulators (TCIs) have repeating unit cell structure, a modular approach was taken to design their circuit implementation.  
Modular circuit boards were designed to have four resonators, each of which is composed of a two 200-pF capacitors and one 1-$\mu$H inductor arranged in a pi-network configuration. The fundamental resonance frequency of each resonator was designed to be 15 MHz and was measured to be 15 $\pm$ 0.5 MHz due to the tolerance of individual passive components. 
Each resonator supports up to six connections (four in-plane and two out-of-plane), which are enabled by reconfigurable solder jumpers. Each in-plane connection additionally has the option of positive or negative coupling. Strong and weak coupling were realized by connecting a either a 100-pF capacitor or a 10-pF capacitor between two resonators in series. 
Arrays of the modular circuit boards were electrically connected to construct the 2D quadrupolar and 3D octupolar TCIs. Periodic boundaries are enforced by connecting the appropriate boundary resonators together (see Fig~\ref{fig:periodicBC}). 

\vspace{12pt}

\noindent
\textit{Spectral measurements:} 
One-port reflection ($S_{11}$) measurements were taken at each resonator using a vector network analyzer (Keysight E5063A) to obtain the spectral density of states (DOS), similar to the procedure in \cite{peterson2018quad,peterson2020FCA,Peterson2021_discl}. The probe used for these measurements was a 50-$\Omega$ SMA connector that was locally grounded to the device under test and terminated with a 24-pF capacitor. The power absorptance, which is defined as the ratio of absorbed to incident power, was calculated as $A(f) = 1-|{S_{11}(f)}^2|$. From the power absorptance, we calculated the DOS $D(f)$ for each resonator as $D(f) = A(f)/f^2$, which accounts for stronger coupling to our capacitive probe at higher frequencies. Since each resonator supports a single mode within the measured spectrum, we normalized D(f) such that the local density of states measured at each individual resonator across all bands integrates to 1. It follows that for a $N$-resonator system, there are $N$ modes in total.

We note that there is some band gap misalignment between resonators that support different spectral features, e.g. resonators at the edges of our material compared to resonators in the bulk. This misalignment is a consequence of both uncompensated loading effects and disorder in our system. Loading effects in the metamaterial can be understood by considering the systematic variation in the local capacitive load at different resonators based on their connectivity. For example, in our quadrupolar TCI a bulk resonator experiences a greater capacitive load than an edge resonator because it is strongly coupled to a greater number of neighbors. This increased capacitive loading cause the bulk bands to shift downward in frequency, relative to the edge bands. 
In addition, the implementation of negative coupling results in a systemic asymmetry of the positive and negative coupling rates.
There is also inherent disorder in the system, which is attributed to the manufacturing tolerances of the discrete components used to construct the resonators.

\vspace{24pt}

% \newpage

\bibliographystyle{naturemag}
{\footnotesize 
    \bibliography{ref}
}

\vspace{24pt}

%%%%%%%%%%%%%%%%%%%%%%%%%%%%%%%%%%%%%
\section*{Acknowledgments}
%%%%%%%%%%%%%%%%%%%%%%%%%%%%%%%%%%%%%

This work was sponsored by the Multidisciplinary University Research Initiative (MURI) grant N00014-20-1-2325 and the US National Science Foundation EFRI grant EFMA-1641084. 
G.B. would additionally like to acknowledge support from the Office of Naval Research (ONR) Director for Research Early Career grant N00014-17-1-2209, and the Presidential Early Career Award for Scientists and Engineers. 
S.S.Y. would also like to acknowledge support from the US National Science Foundation Graduate Research Fellowship Program under Grant No. DGE – 1746047.
%

%%%%%%%%%%%%%%%%%%%%%%%%%%%%%%%%%%%%%
\section*{Author contributions}
%%%%%%%%%%%%%%%%%%%%%%%%%%%%%%%%%%%%%

S.S.Y simulated and designed the electronic circuits, performed experimental measurements, and produced the experimental figures. T.L. and M.L. guided the topological insulator design and performed theoretical calculations. C.W.P. provided guidance about the circuit design and measurement technique. T.L.H. and G.B. supervised all aspects of the project. All authors jointly wrote the paper.

%%%%%%%%%%%%%%%%%%%%%%%%%%%%%%%%%%%%%
\section*{Data availability}
%%%%%%%%%%%%%%%%%%%%%%%%%%%%%%%%%%%%%

The data that support the findings of this study are available from the corresponding author upon reasonable request.

\newpage

\newcommand{\beginsupplement}{%
        \setcounter{table}{0}
        \renewcommand{\thetable}{S\arabic{table}}%
        \setcounter{figure}{0}
        \renewcommand{\thefigure}{S\arabic{figure}}%
        \setcounter{equation}{0}
        \renewcommand{\theequation}{S\arabic{equation}}%\
        \setcounter{section}{0}
        \renewcommand{\thesection}{S\arabic{section}}%
}

\beginsupplement

\begin{center}
\Large{\textbf{Supplementary Figures:}}

\Large{\textbf{Bound states at partial dislocation defects in multipole higher-order topological insulators}
} \\
\vspace{12pt}
\vspace{12pt}
\large{{Sasha S. Yamada}$^1$,
{Tianhe Li}$^2$,
{Mao Lin}$^2$,
{Christopher W. Peterson}$^1$,
{Taylor L. Hughes}$^2$,
and {Gaurav Bahl}}$^3$ \\
\vspace{12pt}
    {\footnotesize{$^1$Department of Electrical and Computer Engineering,}} \\
    {\footnotesize{$^2$Department of Physics and Institute for Condensed Matter Theory,}} \\
    {\footnotesize{$^3$Department of Mechanical Science and Engineering, \\ University of Illinois at Urbana-Champaign, Urbana, IL, USA}}\\
\end{center}

\begin{figure}[!htp]
    \begin{adjustwidth}{-1in}{-1in}
    \centering
    \includegraphics[width=1.4\textwidth]{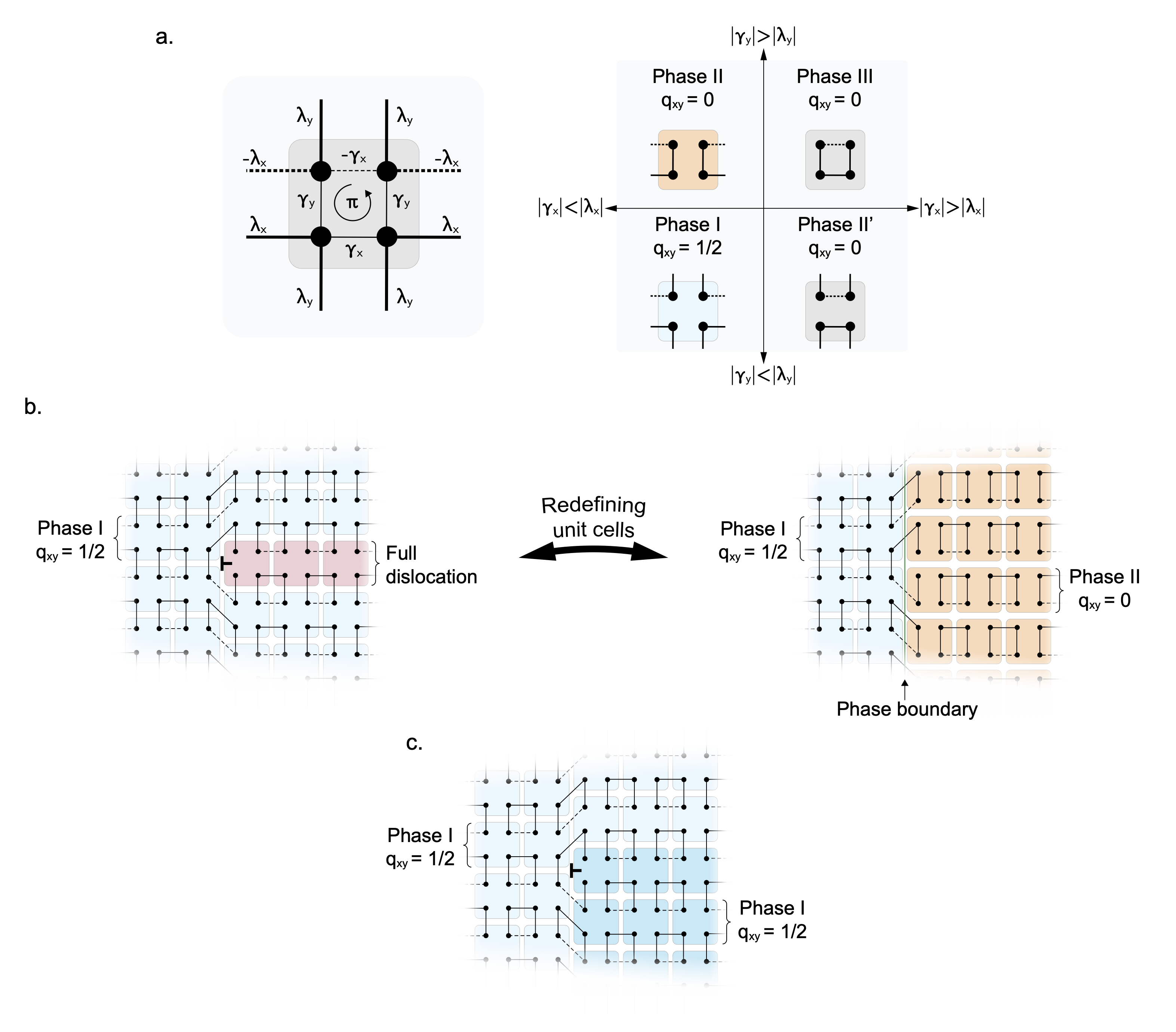}
    \caption{
    \textbf{Showing that full dislocations are featureless at defect core.}
     \textbf{(a)} Possible connections in the minimal model of a single quadrupolar topological crystalline insulator (TCI) unit cell and corresponding phase diagram. Intra-cell couplings are labeled as $\gamma_{x,y}$ and inter-cell couplings are labeled as $\lambda_{x,y}$. The dashed lines denote negative coupling. In the phase diagram, the dimerized limit is illustrated, so only the dominant connections are shown.
    \textbf{(b)} 2D quadrupolar TCI (blue unit cells) with inserted full dislocation (red unit cells). Redefining the unit cells of the lattice indeed yields a phase boundary, but does not trap any 0D bound states as there is no higher-order boundary between the two phases. Again, the dimerized limit is illustrated here.
    \textbf{(c)} Attempting to redefine the unit cells in a single quadrant yields a trivial result, as both regions are in the same topological phase.
    }
    \label{fig:full_disloc}
    \end{adjustwidth}
\end{figure}

\newpage

\begin{figure}[!htp]
    \begin{adjustwidth}{-1in}{-1in}
    \centering
    \includegraphics[width=1.2\textwidth]{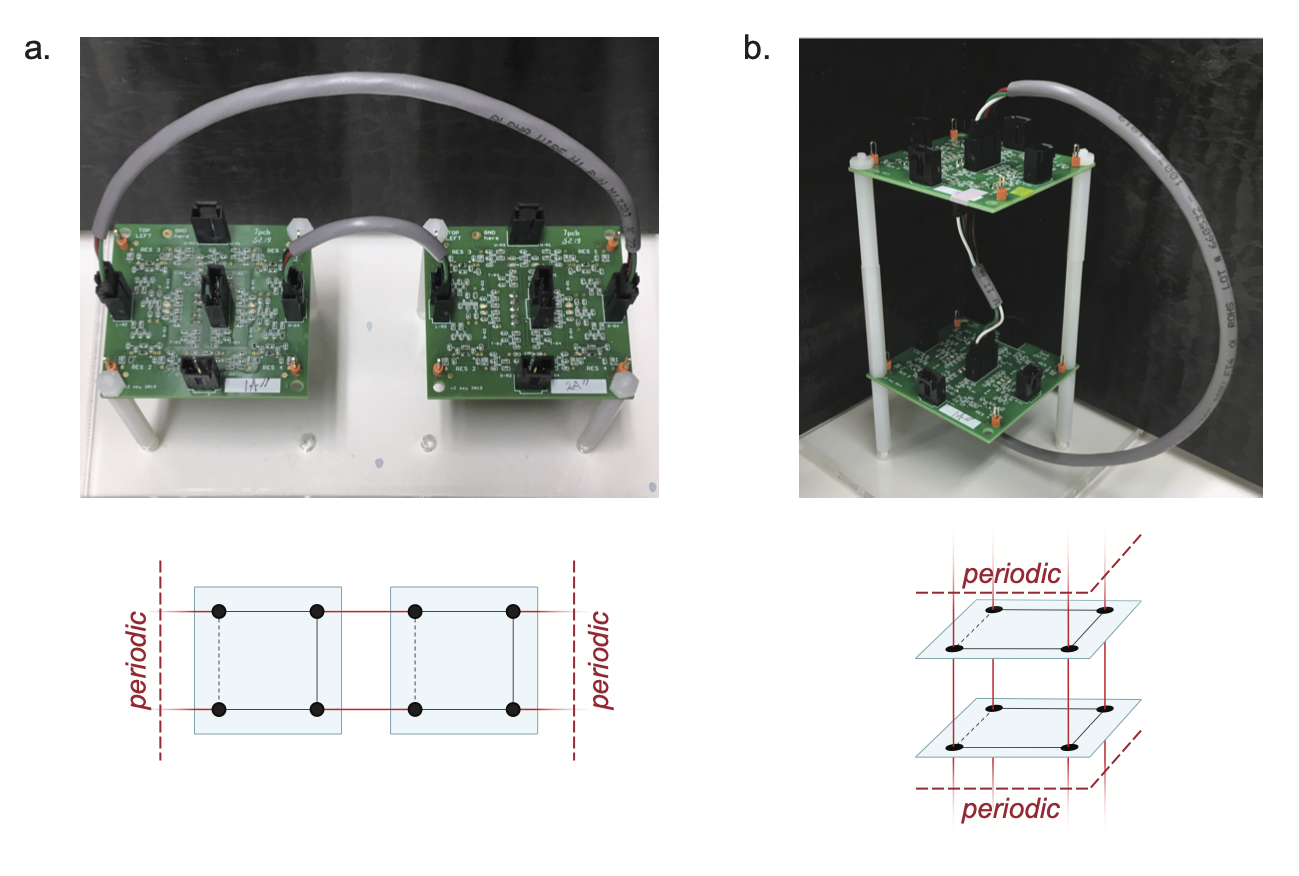}
    \caption{
    \textbf{Implementing periodic boundary conditions.}
    \textbf{(a)} Example of in-plane and
    \textbf{(b)} out-of-plane periodic boundary conditions using two four-site unit cells. Periodicity is enforced by physically wiring the two unit cells together.
    }
    \label{fig:periodicBC}
    \end{adjustwidth}
\end{figure}

\newpage

\begin{figure}[!htp]
    \begin{adjustwidth}{-1in}{-1in}
    \centering
    \includegraphics[width=1.2\textwidth]{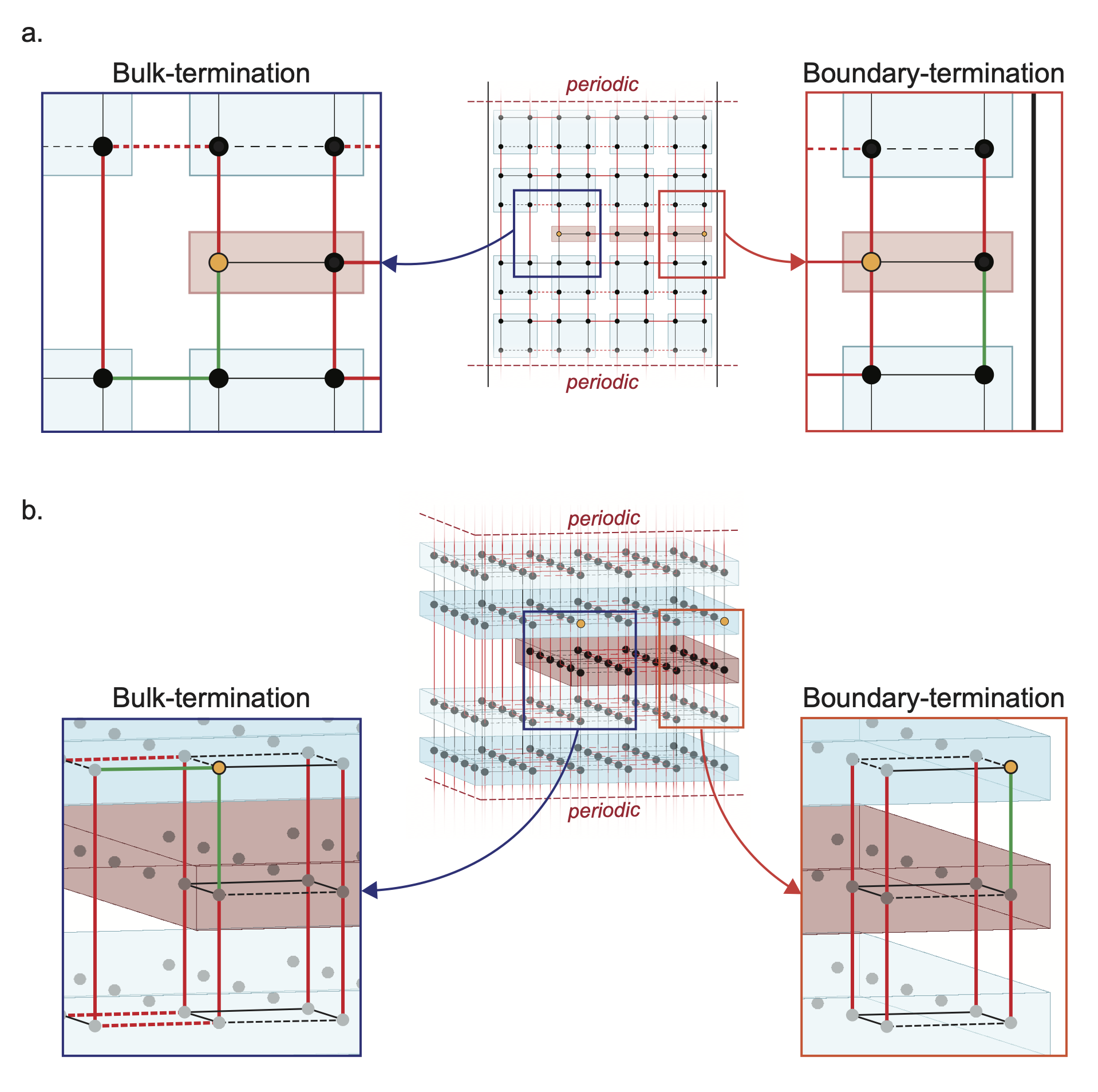}
    \caption{
    \textbf{Local deformation of coupling to spatially isolate bound states to single sites.}
    \textbf{(a)} Representative examples of how the coupling pattern was locally deformed to spatially isolate the bound states associated with bulk- and boundary-terminated partial dislocation defects to a single site in 2D quadrupolar topological crystalline insulators (TCIs).
    The red (black) lines denote strong (weak) coupling, and the dashed lines denote negative coupling. The green lines correspond to where the coupling strength was changed from strong to weak coupling in order to isolate a bound state on the site highlighted in yellow.
    \textbf{(b)} same as (b) but for partial dislocation plane defects in 3D octupolar TCIs. 
    %\GBcomment{can we zoom out a bit? Takes much effort to understand what we're looking at}
    }
    \label{fig:3Dcoupling}
    \end{adjustwidth}
\end{figure}

\begin{figure}[!htb]
    \begin{adjustwidth}{-1in}{-1in}
    \centering
    \includegraphics[width=1.4\textwidth]{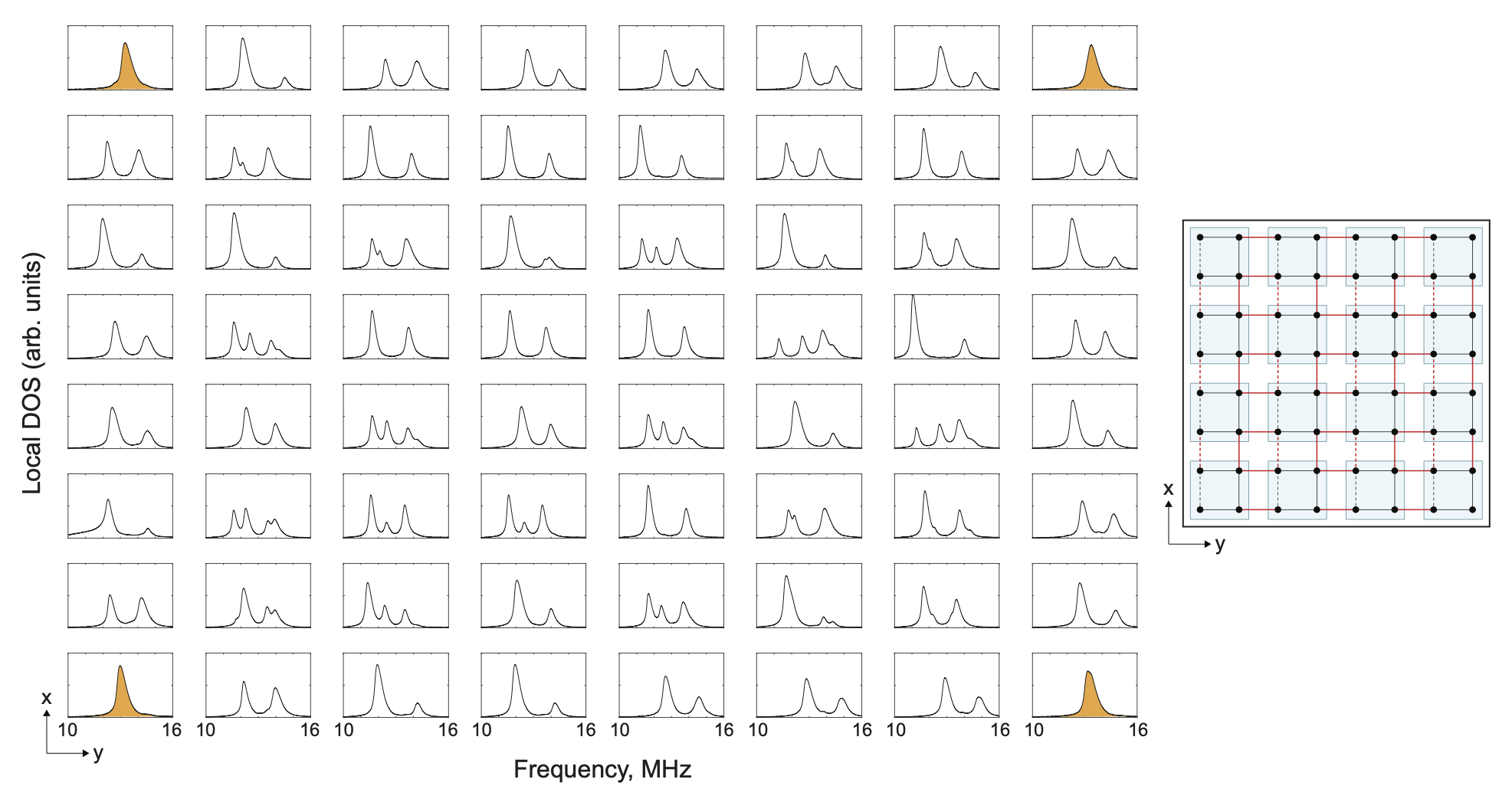}
    \caption{
    \textbf{Measured local DOS at each site for a $4 \times 4$ unit-cell quadrupolar TCI.} 
    These measurements correspond to main manuscript Fig.~\ref{fig:implementation}c. 
    The organization of the grid follows the topology on the right, where the red (black) lines denote strong (weak) coupling, and the dashed lines denote negative coupling.
    In the DOS measurements, gapless modes are highlighted in yellow, and all other sites host gapped modes. 
    }
    \label{fig:quad_grid1}
    \end{adjustwidth}
\end{figure}

\begin{figure}
    \begin{adjustwidth}{-1in}{-1in}
    \centering
    \includegraphics[width=1.4\textwidth]{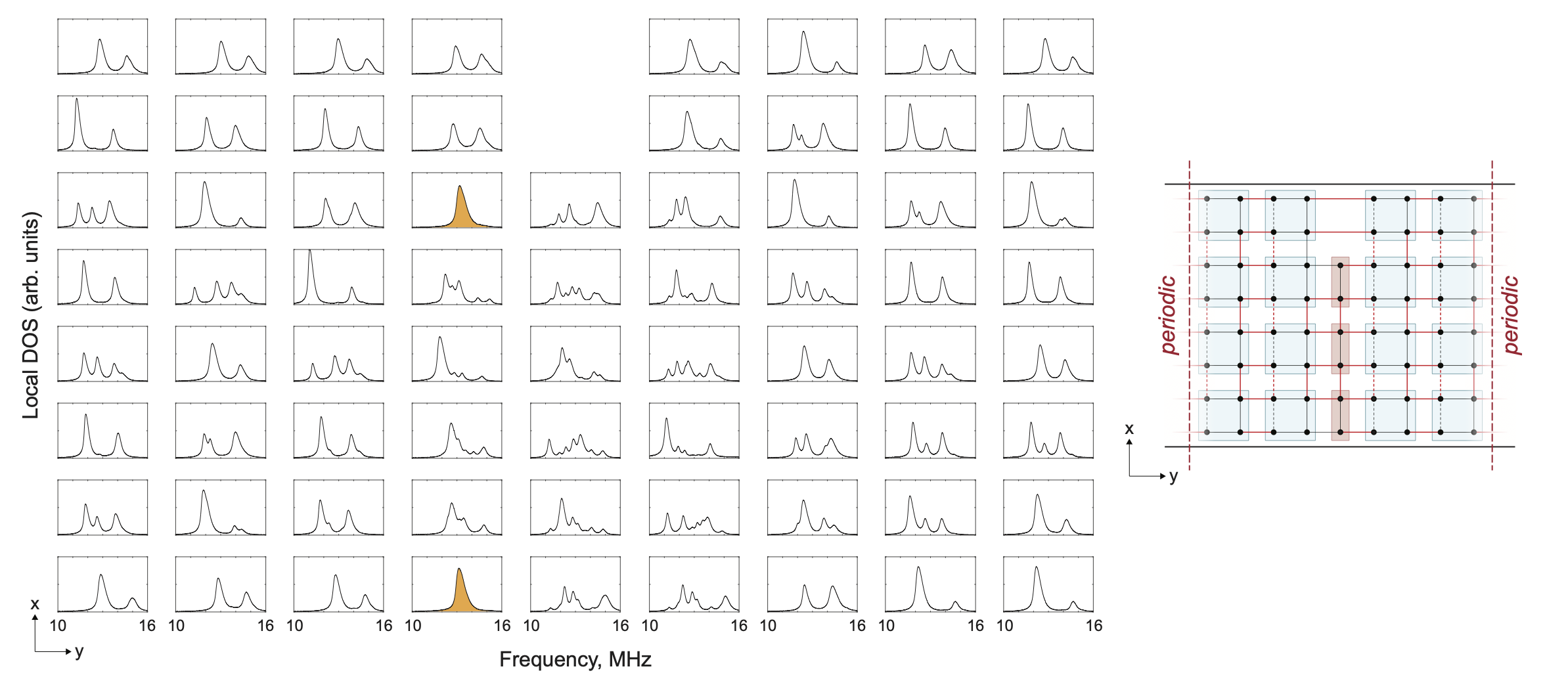}
    \caption{
    \textbf{Measured local DOS at each site for a $4 \times 4$ unit-cell quadrupolar TCI with a bulk-terminated partial dislocation.} 
    These measurements correspond to main manuscript Fig.~\ref{fig:quad_results}a. 
    The organization of the grid follows the topology on the right, the red (black) lines denote strong (weak) coupling, and the dashed lines denote negative coupling. In the DOS measurements, gapless modes are highlighted in yellow, and all other sites host gapped modes. 
    }
    \label{fig:quad_grid2}
    \end{adjustwidth}
\end{figure}

\begin{figure}
    \begin{adjustwidth}{-1in}{-1in}
    \centering
    \includegraphics[width=1.4\textwidth]{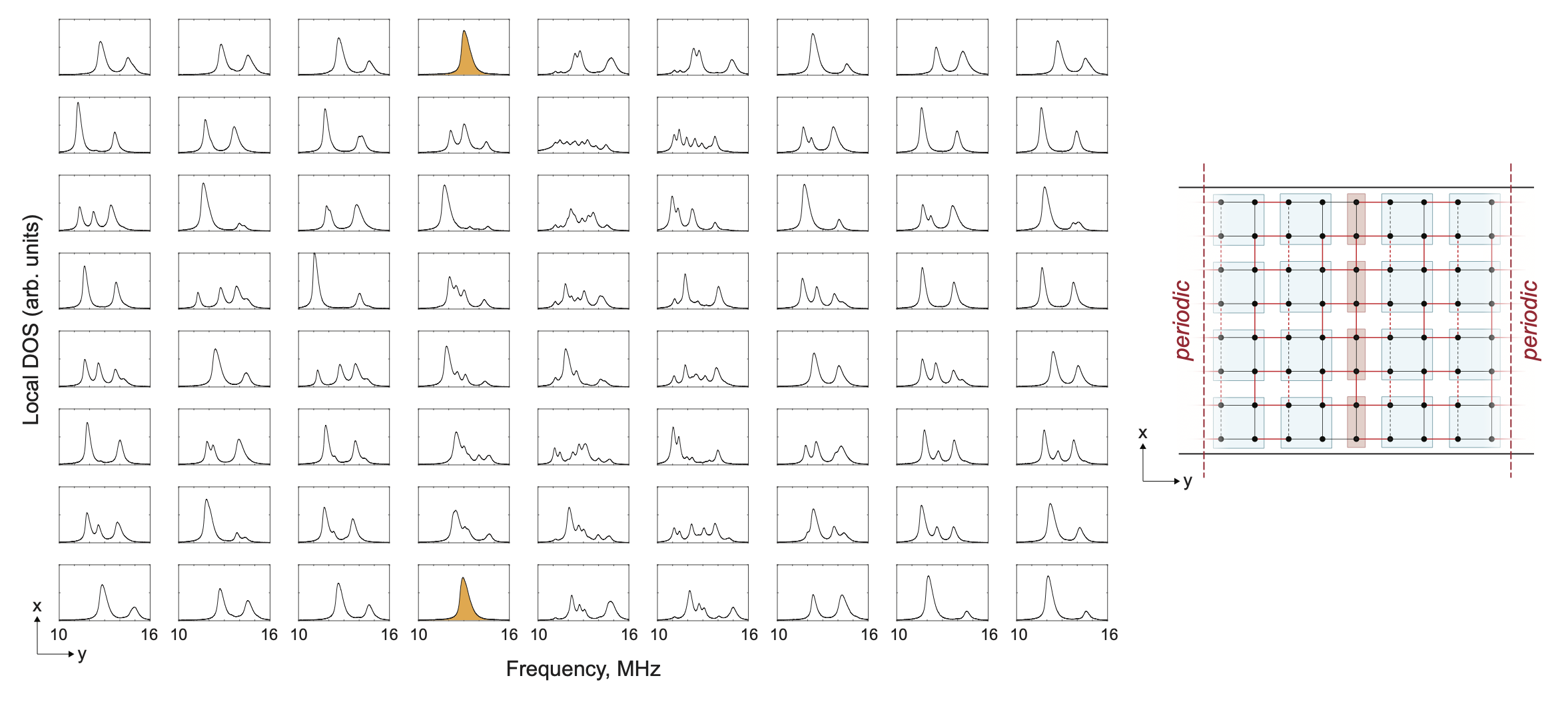}
    \caption{
    \textbf{Measured local DOS at each site for a $4 \times 4$ unit-cell quadrupolar TCI with a boundary-terminated partial dislocation.} 
    These measurements correspond to main manuscript Fig.~\ref{fig:quad_results}b. 
    The organization of the grid follows the topology on the right, the red (black) lines denote strong (weak) coupling, and the dashed lines denote negative coupling. In the DOS measurements, gapless modes are highlighted in yellow, and all other sites host gapped modes. 
    }
    \label{fig:quad_grid3}
    \end{adjustwidth}
\end{figure}

\begin{figure}
    \begin{adjustwidth}{-1in}{-1in}
    \centering
    \includegraphics[width=1.4\textwidth]{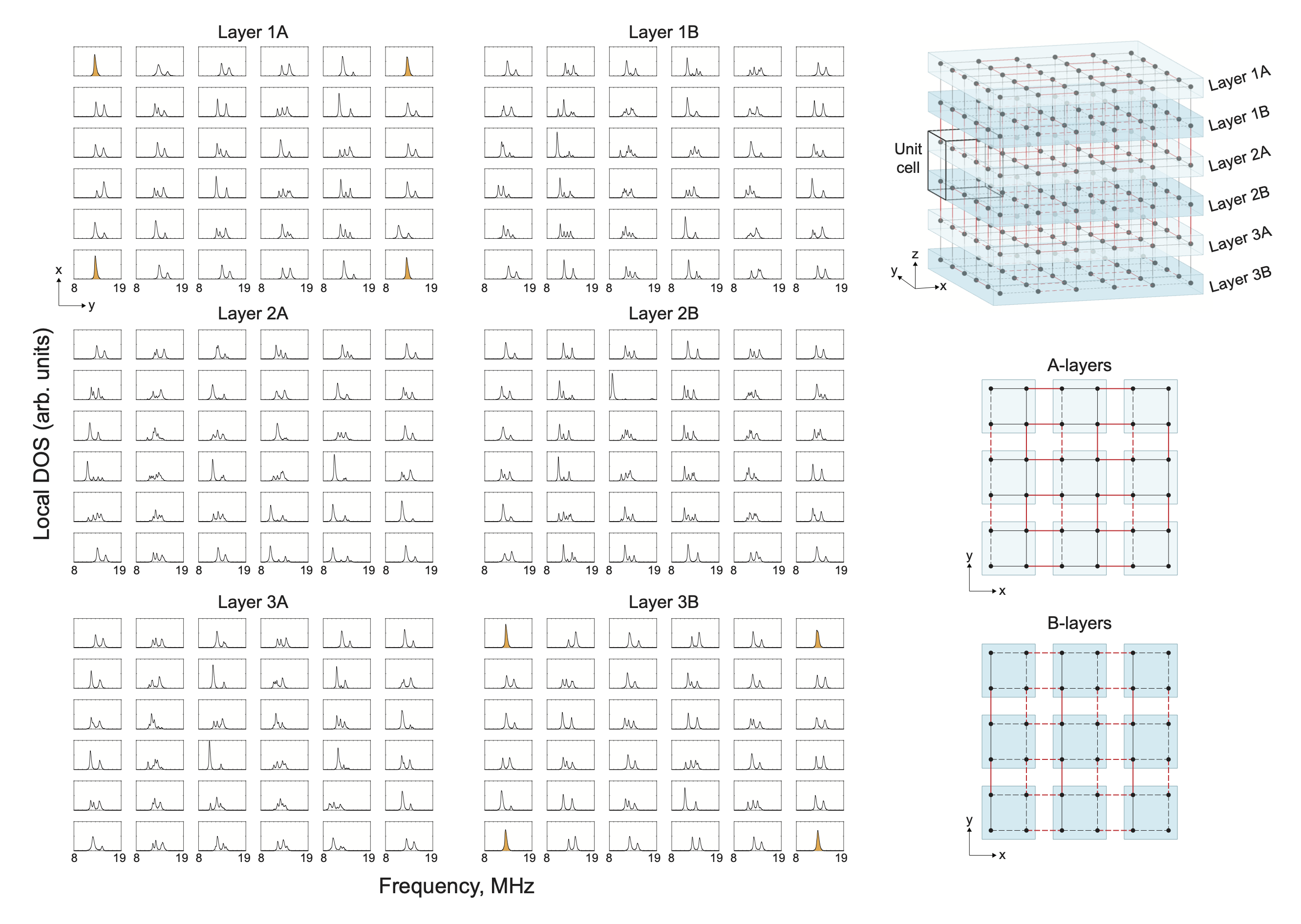}
    \caption{
    \textbf{Measured local DOS at each site for a $3 \times 3 \times 3$ unit-cell octupole TCI with open boundaries.}
    These measurements correspond to main manuscript Fig.~\ref{fig:implementation}f. 
    The organization of the grid follows the topology on the right, the red (black) lines denote strong (weak) coupling, and the dashed lines denote negative coupling. In the DOS measurements, gapless modes are highlighted in yellow, and all other sites host gapped modes. 
    }
    \label{fig:oct_grid1}
    \end{adjustwidth}
\end{figure}

\begin{figure}
    \begin{adjustwidth}{-1in}{-1in}
    \centering
    \includegraphics[width=1.4\textwidth]{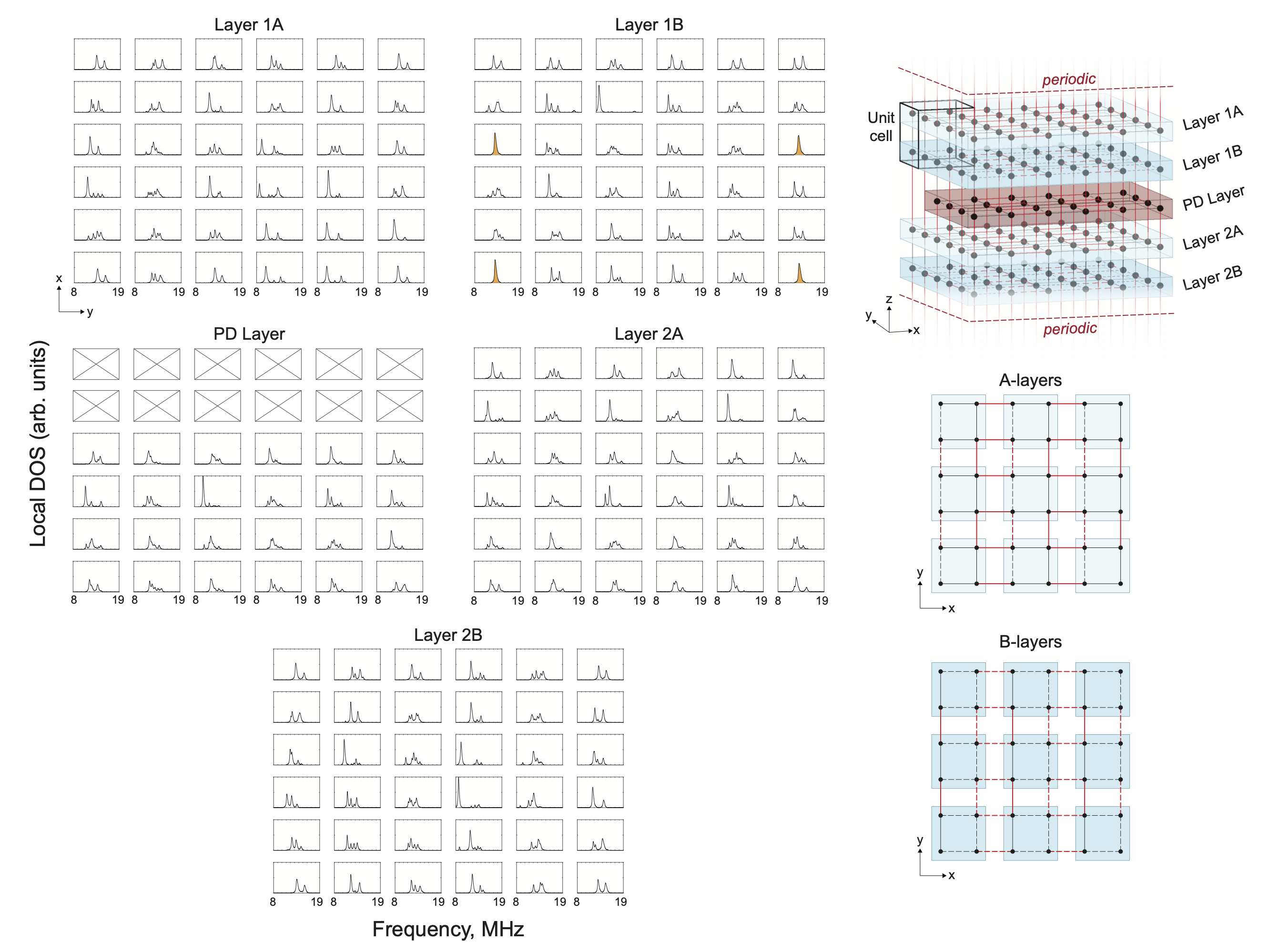}
    \caption{
    \textbf{Measured local DOS at each site for a $3 \times 3 \times 2$ unit-cell octupole TCI with an bulk-terminated partial dislocation.} 
    These measurements correspond to main manuscript Fig.~\ref{fig:oct_results}a.
    The defect is realized as $3 \times 2$ unit-cell `A-layer` quadrupolar TCI.
    The organization of the grid follows the topology on the right, the red (black) lines denote strong (weak) coupling, and the dashed lines denote negative coupling. In the DOS measurements, gapless modes are highlighted in yellow, and all other sites host gapped modes. 
    }
    \label{fig:oct_grid2}
    \end{adjustwidth}
\end{figure}

\begin{figure}
    \begin{adjustwidth}{-1in}{-1in}
    \centering
    \includegraphics[width=1.4\textwidth]{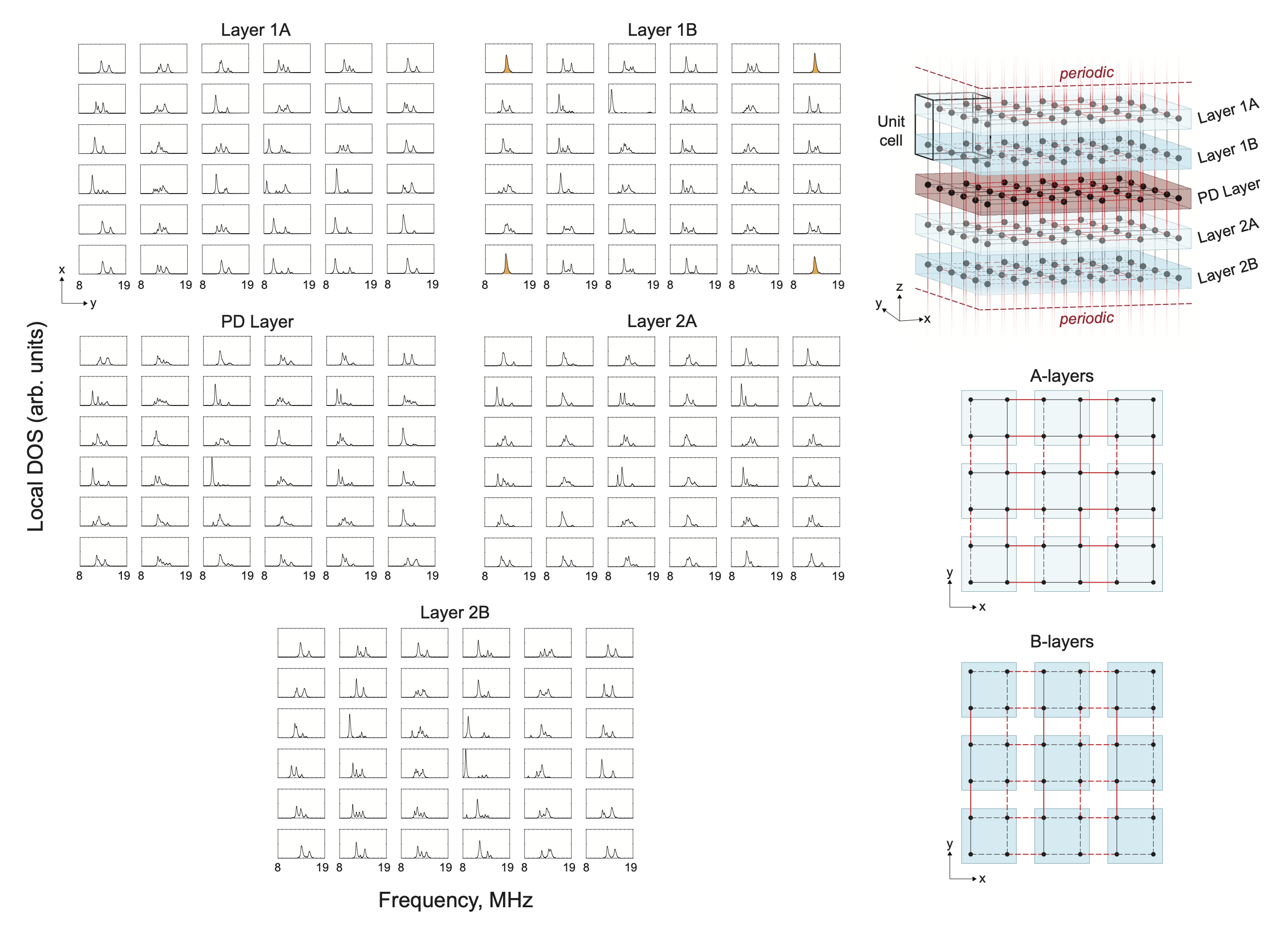}
    \caption{
    \textbf{Measured local DOS at each site for a $3 \times 3 \times 2$ unit-cell octupole TCI with a boundary-terminated partial dislocation.} 
    The defect is realized as an `A-layer` quadrupolar TCI plane. 
    These measurements correspond to main manuscript Fig.~\ref{fig:oct_results}b.
    The organization of the grid follows the topology on the right, the red (black) lines denote strong (weak) coupling, and the dashed lines denote negative coupling. In the DOS measurements, gapless modes are highlighted in yellow, and all other sites host gapped modes. 
    }
    \label{fig:oct_grid3}
    \end{adjustwidth}
\end{figure}

\end{document}